\documentclass{article}
\usepackage[preprint]{spconf}
\usepackage{amsmath}
\usepackage{amssymb}
\usepackage{graphicx}
\usepackage{xcolor} %
\usepackage{booktabs}
\usepackage{multirow}
\usepackage{siunitx}
\usepackage{tikz}
\usepackage[numbers,sort&compress]{natbib}
\usepackage[hidelinks]{hyperref}
\usepackage[capitalize]{cleveref}
\usetikzlibrary{arrows.meta,positioning,calc}

\crefname{figure}{Fig.}{Figs.}
\Crefname{figure}{Fig.}{Figs.}
\crefname{table}{Table}{Tables}
\Crefname{table}{Table}{Tables.}
\crefname{section}{Sec.}{Secs.}
\Crefname{section}{Sec.}{Secs.}

\newcommand{\RR}{\mathbb{R}}
\DeclareMathOperator{\nnz}{nnz}
\DeclareMathOperator{\clip}{clip}

\title{Training-Free Stimulus Encoding for Retinal Implants via Sparse Projected Gradient Descent}
\newif\ifanon
\anonfalse  %

\ifanon
\name{Anonymous ICIP Submission}
\address{}
\else
\name{\begin{tabular}{c}Henning Konermann$^{1}$,
Yuli Wu$^{1,3}$,
Emil Mededovic$^{1}$,
Volkmar Schulz$^{1}$ \\
Peter Walter$^{2}$,
Johannes Stegmaier$^{1,3}$\end{tabular}%
\thanks{This work was supported by Deutsche Forschungsgemeinschaft (DFG, German Research Foundation) with the grant GRK2610: InnoRetVision (project number 424556709).}}
\address{$^{1}$Chair of Imaging and Computer Vision, RWTH Aachen University, Germany \\
         $^{2}$Department of Ophthalmology, RWTH Aachen University, Germany \\
         $^{3}$Faculty of Mathematics and Natural Sciences, Heinrich Heine University Düsseldorf, Germany}
\fi

\toappear{%
  \parbox[t]{\textwidth}{%
    \footnotesize
    \raggedright
    \textcopyright\ 2026 IEEE. Personal use of this material is permitted.
    Permission from IEEE must be obtained for all other uses, in any
    current or future media, including reprinting/republishing this
    material for advertising or promotional purposes, creating new
    collective works, for resale or redistribution to servers or lists,
    or reuse of any copyrighted component of this work in other works.
  }%
}

\begin{document}
\maketitle
\begin{abstract}
Retinal implants aim to restore functional vision despite photoreceptor degeneration,
yet are fundamentally constrained by low resolution electrode arrays and patient-specific perceptual distortions.
Most deployed encoders rely on task-agnostic downsampling and linear brightness-to-amplitude mappings,
which are suboptimal under realistic perceptual models.
While global inverse problems have been formulated as neural networks,
such approaches can be fast at inference,
and can achieve high reconstruction fidelity,
but require training and have limited generalizability to arbitrary inputs.
We cast stimulus encoding as a constrained sparse least-squares problem under a linearized perceptual forward model.
Our key observation is that the resulting perception matrix can be highly sparse,
depending on patient and implant configuration.
Building on this,
we apply an efficient projected residual norm steepest descent solver that exploits sparsity and supports stimulus bounds via projection.
In silico experiments across four simulated patients and implant resolutions from $15\times15$ to $100\times100$ electrodes demonstrate improved reconstruction fidelity,
with up to $+0.265$ SSIM increase,
$+12.4\,\mathrm{dB}$ PSNR,
and $81.4\%$ MAE reduction on Fashion-MNIST compared to Lanczos downsampling.
\end{abstract}
\begin{figure}[h!tb]
  \centering
  \begin{tikzpicture}[
    font=\small,
    block/.style={draw,rounded corners,align=center,minimum width=3.2cm,minimum height=0.95cm},
    smallblock/.style={draw,rounded corners,align=center,minimum width=2.6cm,minimum height=0.8cm},
    arrow/.style={-{Latex[length=2.2mm]},thick},
    dashedarrow/.style={-{Latex[length=2.2mm]},thick,dashed}
  ]
  \node[align=center] (x) {\includegraphics[width=1.35cm]{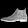}\\\(\mathbf x_t\)};
  \node[block,below=6mm of x] (enc) {Encoder\\(steepest descent step)};

  \coordinate (j) at ($(enc.south)+(0,-10mm)$);
  \node[circle,fill,inner sep=1.2pt] (dot) at (j) {};

  \node[smallblock,right=12mm of dot] (lin) {Linearized model\\\(\hat{\mathbf y}^{(k)}_t=\mathbf P_\phi\,\mathbf s^{(k)}_t\)};
  \node[smallblock,below=10mm of dot] (nonlin) {Nonlinear perception\\\(\mathbf y^{(k)}_t=\mathcal P_\phi(\mathbf s^{(k)}_t)\)};
  \node[align=center,below=7mm of nonlin] (y) {\includegraphics[width=1.35cm]{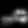}\\\(\mathbf y_t^{(k)}\)};
  \node[smallblock,left=12mm of dot] (delay) {Warm start\\\(\mathbf s^{(0)}_{t+1}\leftarrow \mathbf s^{(K)}_t\)};

  \draw[arrow] (x) -- (enc);
  \draw[dashedarrow] (delay.north) |- node[above,pos=0.5]{\(\mathbf s^{(0)}_t\)} (enc.west);

  \draw[thick](enc) -- node[above right,pos=1]{\(\mathbf s^{(k)}_t\)} (dot);
  \draw[arrow] (dot) -- (lin.west);
  \draw[arrow] (lin.north) |- node[above,pos=0.5]{\(\nabla f(\mathbf s^{(k)}_t)\)} (enc.east);
  \draw[arrow] (dot) -- (nonlin);
  \draw[arrow] (nonlin) -- (y);
  \draw[arrow] (dot) -- (delay.east);
  \end{tikzpicture}
  \caption{Overview of the proposed encoding pipeline.
  The encoder updates the stimulus via projected residual norm steepest descent under the linearized perception model \(\mathbf P_\phi\),
  and produces the percept through the nonlinear forward model \(\mathcal P_\phi\).
  For potential real-time use,
  warm-starting across frames via a one-frame delay ($K=1$) could be employed.}
\label{fig:pipeline}
\end{figure}
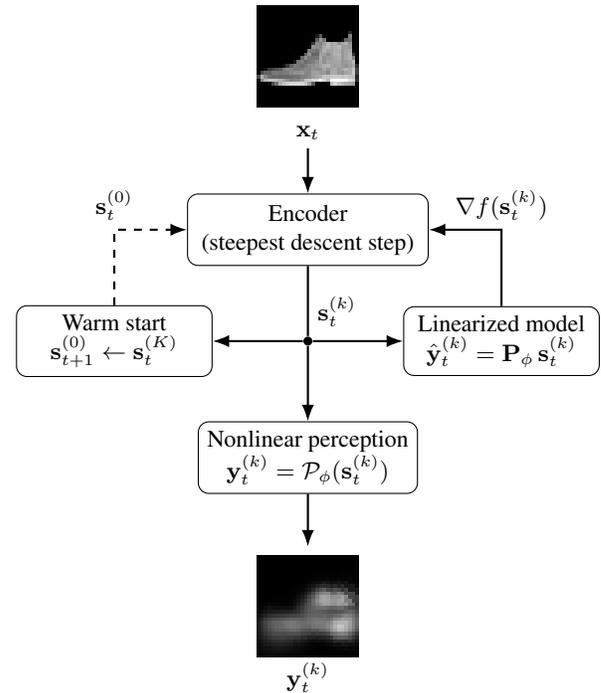

\section{Introduction}
\label{sec:intro}
Retinal prostheses restore rudimentary vision by electrically stimulating surviving retinal neurons via e.g. epiretinal or subretinal electrode arrays.
Clinical systems such as Argus~II~\cite{Luo2016} employ 60 electrodes,
while more recent devices like PRIMA~\cite{Holz} feature 378 electrodes.
Future prototypes target thousands of electrodes \cite{Bhuckory2025},
increasing computational and memory pressure for stimulus encoding.
Current clinical encoders largely downsample the scene to the electrode grid,
and map intensity to amplitudes under hardware constraints \cite{Luo2016}.
While high-quality downsampling filters can improve performance in human testing \cite{Barnes2016},
other work also considers additional preprocessing such as contrast enhancement or task-specific feature amplification \cite{Goldstein2025},
purely filter-based approaches ignore distortions induced by axonal activation and electrode-dependent perceptual spread,
which are captured by modern perceptual models \cite{Granley2021,Beyeler2017}.
Conversely,
learning-based encoders can optimize end-to-end metrics,
but typically require retraining and may generalize poorly to arbitrary inputs,
raising practical deployment and safety concerns \cite{Downing2025}.

This paper targets \emph{efficient,
scalable,
and bound-aware} stimulus optimization under a linearized perceptual model,
by exploiting the inherent sparse nature of the problem.
Importantly,
iterative solvers provide an anytime speed--quality trade-off and enable warm-starting across frames,
making real-time encoding feasible on modern hardware.
Our contributions are:
(i) an empirical characterization showing that the linearized perception model can exhibit extreme sparsity,
with simulated number of non-zeros (nnz) below $1\%$,
depending on patient and implant configuration;
(ii) a sparse,
bound-aware projected residual norm steepest descent solver that exploits this structure via sparse matrix--vector products and simple feasibility projections;
and (iii) an \emph{in silico} evaluation across four simulated patients and implant sizes,
showing up to $+0.265$ SSIM increase,
$+12.4\,\mathrm{dB}$ PSNR,
and $81.4\%$ MAE reduction on F-MNIST compared to Lanczos.

\section{Related work}
\label{sec:rw}
Prior work spans \textbf{classical} approaches that downsample to the electrode grid with linear intensity-to-amplitude mapping,
often with unspecified downsampling filters \cite{Luo2016,Barnes2016},
\textbf{perceptual models} that simulate phosphenes using axon maps and current spread \cite{Beyeler2017,Granley2021,Granley2023},
\textbf{learning-based} methods that train differentiable encoders end-to-end under simulators \cite{RuytervanSteveninck2022,Wu2023,Wu2024,Granley2022,Granley2023},
and \textbf{linearized inverse} methods that use pseudo-inverse least squares under linearized perception with feasibility handled post hoc \cite{Fauvel2022}.
In this paper,
we focus on \emph{training-free} encoding,
and treat learning-based encoders as complementary approaches.
Specifically,
we study sparse,
bound-aware iterative solvers that scale to larger implants,
and provide an anytime trade-off for real-time encoding.

\section{Problem formulation}
\label{sec:form}
We consider grayscale input images $\mathbf x \in [0,1]^{M}$ (vectorized) representing a desired percept on an $M$-pixel simulation grid.
We denote normalized stimulation patterns by $\mathbf s \in [0,1]^{N}$,
where $N$ is the number of electrodes.
We model perception using a patient-specific \emph{perception} function:
\begin{equation}
\mathbf y = \mathcal{P}_{\phi}(\mathbf s),
\label{eq:forward}
\end{equation}
where $\mathcal{P}_{\phi}:[0,1]^N \rightarrow [0,1]^{M}$ maps electrode-space stimuli to simulated percept pixels,
and $\mathbf y \in [0,1]^{M}$ is the (generally nonlinear) predicted percept,
and $\phi$ denotes patient and implant parameters (e.g.,
axon map geometry) \cite{Granley2021,Beyeler2017,Granley2023}.
Importantly,
$\phi$ can be adapted to patient-specific effects that deviate from standard models.
We use elementwise normalized stimuli $\mathbf s \in [0,1]^N$,
since physical scaling can be absorbed into $\mathcal{P}_{\phi}$.

\textbf{Linear perception matrix and sparsity.}
To obtain an efficient solver,
we use a linearized perception model:
\begin{equation}
\mathbf y_{\mathrm{lin}} = \mathbf P_{\phi}\,\mathbf s,
\label{eq:lin}
\end{equation}
where $\mathbf P_{\phi} \in \RR^{M \times N}$ is a \emph{sparse} perception matrix.
In this paper,
we assume $\mathbf P_{\phi}$ is given (precomputed for a fixed patient and implant),
and focus on efficient inversion; estimation and calibration of patient-specific parameters are outside our scope (see e.g.
\cite{Fauvel2022,Granley2023}).
This linear model can either be assumed directly,
or obtained empirically by activating electrodes individually and recording the simulated percept,
such that column $n$ of $\mathbf P_{\phi}$ corresponds to the percept elicited by electrode $n$ under a unit stimulus \cite{Fauvel2022}.
More generally,
$\mathbf P_{\phi}$ can also be obtained by locally linearizing $\mathcal{P}_{\phi}$ around a working point defined by the current stimulus and its corresponding percept.
Crucially,
$\mathbf P_{\phi}$ is typically sparse because each electrode affects only a localized region of the percept.

\begin{figure}[!t]
  \centering
  \includegraphics[width=\columnwidth]{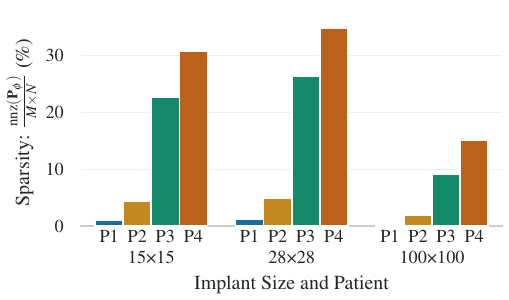}
  \caption{Sparsity of the perception matrix $\mathbf P_{\phi}$ across implant resolutions and patient configurations.
  Percentages are computed after thresholding values below 5\% of the maximum matrix brightness. Patients as in~\cref{sec:exp}.
  Additional experiments analyzing the effect of sparsity truncation are provided in the supplementary material.}
  \label{fig:perception_matrix_sparsity}
\end{figure}

As shown in \cref{fig:perception_matrix_sparsity},
the effective sparsity of $\mathbf P_{\phi}$ varies substantially across patient configurations and implant sizes,
in particular for large-area implants,
and for settings where $\rho$ and $\lambda$ are small relative to the overall implant dimensions,
which motivates using sparse matrix formats and sparse matrix--vector products throughout our solver.

\textbf{Stimulus encoding} seeks a feasible stimulus whose predicted percept matches the target,
i.e.,
we aim to solve a constrained nonlinear inverse problem,
which we approximate using the linearized model defined in \cref{eq:lin}:
\begin{equation}
\arg\min_{\mathbf s \in [0,1]^N} \left\| \mathcal{P}_{\phi}(\mathbf s) - \mathbf x \right\|_2^2
\;\approx\;
\arg\min_{\mathbf s \in [0,1]^N} \left\| \mathbf P_{\phi}\mathbf s - \mathbf x \right\|_2^2.
\label{eq:cls}
\end{equation}
An ablation comparing evaluation under the linearized and nonlinear perceptual model is provided in the supplementary material.
\section{Proposed method}
\label{sec:method}
A closed-form pseudo-inverse solution to the unconstrained problem,
as considered in \cite{Fauvel2022},
is $\mathbf s^\star = \mathbf P_\phi^{+}\mathbf x$.
However,
this baseline is ill-suited for practical stimulus encoding under \cref{eq:cls}.
First,
it does not enforce feasibility and can produce negative or overly large amplitudes,
so additional post hoc clipping and scaling is required.
Second,
the pseudo-inverse is generally dense,
so explicitly forming or storing $\mathbf P_\phi^{+}\in\mathbb{R}^{N\times M}$ becomes prohibitive for high-resolution implants.

\begin{figure}[!t]
  \centering
  \includegraphics[width=\columnwidth]{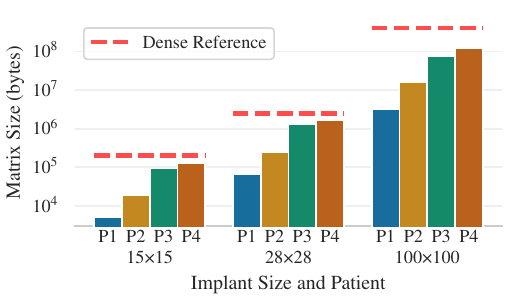}
  \caption{Sparse matrix size across electrode counts,
  shown for $N=M$ equal to the electrode count.
  Truncation is performed as in \cref{fig:perception_matrix_sparsity} and patients as in \cref{sec:exp}.}
  \label{fig:matrix_size}
\end{figure}

\textbf{Memory footprint.}
A dense pseudo-inverse has $NM$ entries,
and becomes prohibitive at high resolutions,
see \cref{fig:matrix_size}.
Instead,
we store $\mathbf P_\phi$ in sparse CSR format \cite{Saad2003}.
Storage scales as
\begin{equation}
\mathcal{O}(\nnz(\mathbf P_\phi)+M+N).
\end{equation}
\textbf{Warm-started projected residual norm steepest descent for real-time encoding.}
To avoid explicitly forming $\mathbf P_\phi^{+}$ while enforcing feasibility,
we solve \cref{eq:cls} via projected residual norm steepest descent \cite{Saad2003} on
\begin{equation}
f(\mathbf s)=\tfrac12\left\|\mathbf P_\phi\mathbf s-\mathbf x\right\|_2^2.
\end{equation}
Each iteration only requires sparse matrix--vector products with $\mathbf P_\phi$ and $\mathbf P_\phi^\top$;
runtime can be further improved by caching $\mathbf P_\phi^\top$.
An overview of the resulting encoding loop is shown in \cref{fig:pipeline}.

For \cref{eq:cls},
the gradient is
$\nabla f(\mathbf s)= \mathbf P_\phi^\top(\mathbf P_\phi\mathbf s-\mathbf x)$.
Residual norm steepest descent performs iterations indexed by $k$:
\begin{equation}
\mathbf s^{(k+1)} = \clip_{[0, 1]}\!\left(\mathbf s^{(k)} - \alpha^{(k)} \nabla f(\mathbf s^{(k)})\right),
\label{eq:pgd}
\end{equation}
where $\alpha^{(k)}$ is an exact line-search step size for the least-squares objective \cite{Saad2003},
i.e.,
\begin{equation}
\alpha^{(k)}=\frac{\left\|\nabla f(\mathbf s^{(k)})\right\|_2^2}{\left\|\mathbf P_\phi \nabla f(\mathbf s^{(k)})\right\|_2^2},
\end{equation}
and $\clip_{[0,1]}$ denotes elementwise clipping,
i.e.,
$\clip_{[0,1]}(\mathbf z)=\min\!\left(1,\max\!\left(0,\mathbf z\right)\right)$,
with $\min$ and $\max$ applied elementwise.
For potential real-time use with sequential inputs indexed by $t$,
warm-starting with the previous solution,
$\mathbf s^{(0)}_{t} \leftarrow \mathbf s^{(K)}_{t-1}$,
could reduce iterations needed to reach a target fidelity by exploiting temporal coherence.

\section{Experiments}
\label{sec:exp}
\subsection{Simulation setup}
We evaluate on four simulated patients as specified in the Hybrid Neural Autoencoder setup of \cite{Granley2022},
using the pulse2percept simulation framework \cite{Beyeler2017} with the perceptual model of \cite{Beyeler2019} as in \cite{Fauvel2022}.
Patients are defined by the axon map parameters \(\rho\) and \(\lambda\) (cf.\
\cite{Granley2022}),
where \(\rho\) controls percept size and \(\lambda\) controls eccentricity.
An additional robustness experiment analyzing patient model mismatch by misspecifying \(\rho\) and \(\lambda\) is provided in the supplementary material.

We consider three implant resolutions,
$15{\times}15$ (225 electrodes,
$400/400\,\si{\micro\metre}$ spacing,
as in \cite{Granley2022}),
$28{\times}28$ (784 electrodes,
$200/200\,\si{\micro\metre}$ spacing,
matching MNIST resolution),
and $100{\times}100$ (10{,}000 electrodes,
$100/100\,\si{\micro\metre}$ spacing,
a potential future large-scale implant).

For each implant,
we generate $\mathbf P_\phi$ as in \cite{Fauvel2022} by activating each electrode individually and recording the elicited percept,
then optimize $\mathbf s$ per input image.

Unless stated otherwise,
we use standard pulse2percept parameters following \cite{Fauvel2022}.
Full simulation parameters and implant configurations are listed in the supplementary material.

\subsection{Baselines}
\label{sec:baselines}
We compare against:
\begin{itemize}
  \item Nearest-neighbor and Lanczos downsampling \cite{Barnes2016}.
  \item Pseudo-inverse least-squares solution under the linearized model with post hoc clipping and scaling \cite{Fauvel2022}.
\end{itemize}

\noindent\textbf{Note on baseline availability.}
Non-learning baselines for prosthetic stimulus encoding are scarce.
We therefore compare against standard downsampling and a reproducible pseudo-inverse baseline under the linearized model,
using standard feasibility operations as in \cite{Fauvel2022}.
We include Lanczos downsampling because \cite{Barnes2016} is one of the few works we found that explicitly specifies the downsampling filter,
whereas many related papers do not report the concrete method.

For a fair comparison,
we allow nearest-neighbor and Lanczos baselines to apply a global rescaling of the resulting stimulus vector \(\mathbf s\).
Using the linearized perceptual model \(\mathbf y=\mathbf P_\phi \mathbf s\),
we set \(\alpha = \max(\mathbf x) / \max(\mathbf P_\phi \mathbf s)\) and replace \(\mathbf s \leftarrow \clip_{[0,\mathbf 1]}(\alpha\,\mathbf s)\),
which adjusts the stimulus to match the maximum expected brightness of the target percept \(\mathbf x\) under \(\mathbf P_\phi\).

For the pseudo-inverse baseline,
we compute an unconstrained stimulus \(\tilde{\mathbf s}=\mathbf P_\phi^{+}\mathbf x\),
clip negative amplitudes (\(\tilde{\mathbf s}\leftarrow \max(\mathbf 0,\tilde{\mathbf s})\)),
and then apply the same maximum-brightness scaling under \(\mathbf y=\mathbf P_\phi \tilde{\mathbf s}\) as above.
Note that \cite{Fauvel2022} evaluates this proof-of-concept baseline under the linearized model,
which can yield infeasible stimuli without explicit bound handling.

\subsection{Data and metrics}
We evaluate on MNIST \cite{Lecun1998},
Fashion-MNIST (F-MNIST) \cite{Xiao2017},
and the COCO Panoptic dataset \cite{Lin2014}.
For MNIST and F-MNIST,
we use the standard test sets,
with 10{,}000 images each.
For COCO,
we follow the preprocessing procedure of \cite{Granley2022},
and convert images to grayscale using \cite{Lu2014}.
We use the validation split,
totalling 1{,}974 images in our COCO subset.
Metrics are computed between predicted nonlinear percept $\mathbf y = \mathcal{P}_{\phi}(\mathbf s)$ and target $\mathbf x$:
SSIM~\cite{Wang2004},
PSNR,
and MAE.
We report all three metrics in \cref{tab:combined_comparison}.
Quality metrics for residual norm steepest descent are reported after 10{,}000 iterations to ensure convergence.
We use a GPU implementation of \cref{sec:method} using CuPy \cite{Okuta2017} on a NVIDIA GTX 1660.

\section{Results}
\label{sec:results}

\begin{figure*}[!t]
  \centering
  \includegraphics[width=\textwidth]{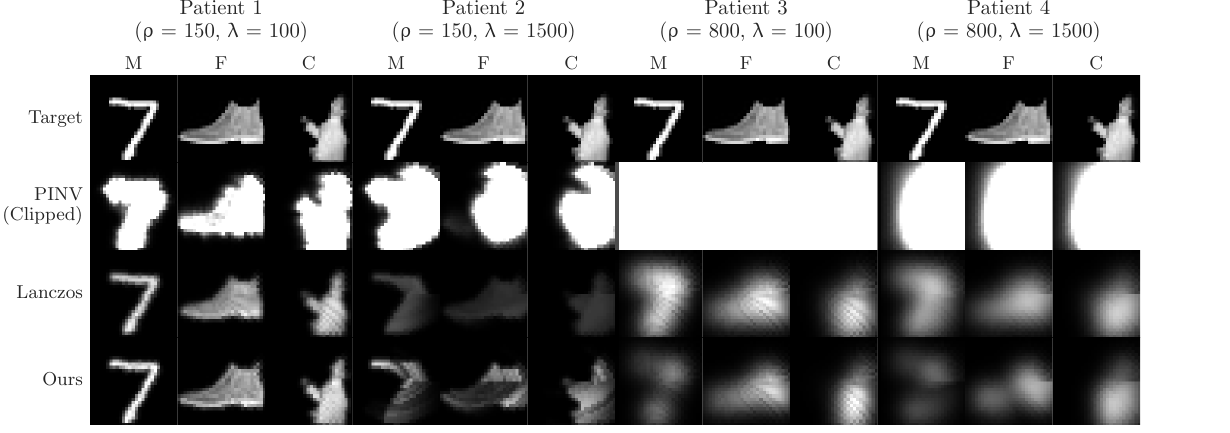}
  \caption{Qualitative comparison across all simulated patients and datasets (M: MNIST,
  F: F-MNIST,
  C: COCO).
  Rows show the target percept,
  Lanczos,
  pseudo-inverse with post hoc clipping (PINV),
  and projected residual norm steepest descent.
  All examples use a $28\times28$ implant and a $28\times28$ simulated perception grid.}
  \label{fig:image_grid_comparison_all_patients}
\end{figure*}

\cref{fig:image_grid_comparison_all_patients} shows qualitative results across all four simulated patients.
For favorable patient parameters,
simple downsampling baselines can already perform satisfactorily (e.g.\
\((\rho=\SI{150}{\micro\metre},
\lambda=\SI{100}{\micro\metre})\)),
but projected residual norm steepest descent produces percepts that more closely match the target,
while the pseudo-inverse baseline often fails due to post hoc feasibility operations,
in particular clipping negative amplitudes (cf.\
\cref{sec:exp}).
Quantitative results are summarized in \cref{tab:combined_comparison} (SSIM/PSNR/MAE between predicted percept \(\mathbf y\) and target \(\mathbf x\)).
Across implant resolutions on F-MNIST,
our method improves over Lanczos for all patients,
with $\Delta$SSIM ranging from $+0.051$ to $+0.265$
across the patient--implant combinations reported.
For the $28\times28$ implant,
$\Delta$SSIM ranges from $+0.051$ to $+0.212$ on F-MNIST,
and from $+0.003$ to $+0.211$ on MNIST.
PSNR and MAE are reported in \cref{tab:combined_comparison}.
On COCO,
high-\(\rho\) patients (\(\rho=\SI{800}{\micro\metre}\)) show $\Delta$SSIM of $+0.129$ and $+0.135$,
while low-\(\rho\) patients (\(\rho=\SI{150}{\micro\metre}\)) show only marginal $\Delta$SSIM ($\le +0.012$),
highlighting strong patient--dataset interactions.

\begin{figure}[!t]
  \centering
  \includegraphics[width=\columnwidth]{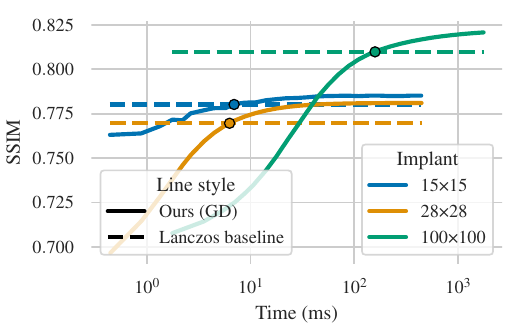}
  \caption{Speed--quality trade-off,
  SSIM versus aggregated iteration time.
  Aggregated time uses GTX1660 per-iteration means
  (\(15\times15\): \SI{0.442}{\milli\second},
  \(28\times28\): \SI{0.445}{\milli\second},
  \(100\times100\): \SI{1.764}{\milli\second}).
  The horizontal line denotes Lanczos SSIM,
  and crossover points mark when our method matches it.}
  \label{fig:precision_vs_time_ssim}
\end{figure}

\cref{fig:precision_vs_time_ssim} shows SSIM as a function of time,
with the Lanczos baseline shown as a reference.
For $15\times15$ electrodes,
our method reaches the Lanczos SSIM value $0.78$ after $t=\SI{6.9}{\milli\second}$,
at 16 iterations.
For $28\times28$ electrodes,
the crossover occurs at SSIM $0.77$ after $t=\SI{6.3}{\milli\second}$,
also at 16 iterations.
For $100\times100$ electrodes,
the crossover occurs at SSIM $0.81$ after $t=\SI{147.1}{\milli\second}$,
at 100 iterations.

\begin{figure}[!t]
  \centering
  \includegraphics[width=\columnwidth]{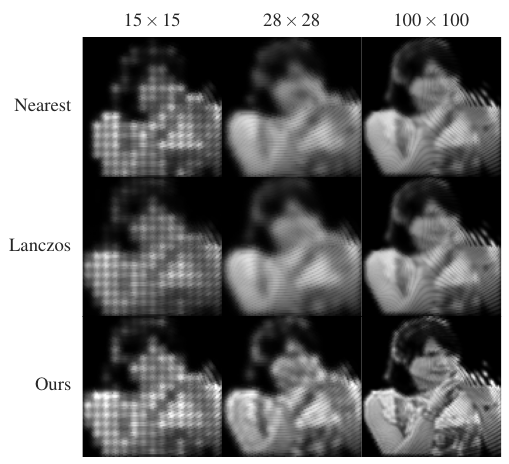}
  \caption{Qualitative comparison across implant resolutions and encoding methods for Patient~1,
  using a $100\times100$ simulated perception grid.}
  \label{fig:implant_quality_patient_1}
\end{figure}

\begin{table*}[!t]
  \centering
  \small
  \begin{tabular}{
    l l c @{\hspace{5pt}} c @{\hspace{5pt}} c @{\hspace{10pt}} c @{\hspace{5pt}} c @{\hspace{5pt}} c @{\hspace{10pt}} c @{\hspace{5pt}} c @{\hspace{5pt}} c @{\hspace{15pt}} c @{\hspace{5pt}} c @{\hspace{5pt}} c @{\hspace{10pt}} c @{\hspace{5pt}} c @{\hspace{5pt}} c @{\hspace{10pt}} c @{\hspace{5pt}} c @{\hspace{5pt}} c
  }
  \toprule%
  & & \multicolumn{9}{c}{Implant Variation} & \multicolumn{9}{c}{Dataset Variation} \\
  \cmidrule(l{10pt}r{20pt}){3-11} \cmidrule(l{5pt}r{10pt}){12-20}
   \multirow{2}{*}{Patient $(\rho,\lambda)$} & \multirow{2}{*}{Method} & \multicolumn{3}{c}{\hspace*{-5pt}$15\times15$} & \multicolumn{3}{c}{\hspace*{-10pt}$28\times28$} & \multicolumn{3}{c}{\hspace*{-15pt}$100\times100$} & \multicolumn{3}{c}{\hspace*{-10pt}MNIST} & \multicolumn{3}{c}{\hspace*{-10pt}F-MNIST} & \multicolumn{3}{c}{\hspace*{-10pt}COCO} \\
   \cmidrule(l{5pt}r{10pt}){3-5}\cmidrule(l{0pt}r{10pt}){6-8}\cmidrule(l{0pt}r{15pt}){9-11}\cmidrule(l{0pt}r{10pt}){12-14}\cmidrule(l{0pt}r{10pt}){15-17}\cmidrule(l{0pt}r{5pt}){18-20}
   & & S$\uparrow$ & P$\uparrow$ & M$\downarrow$ & S$\uparrow$ & P$\uparrow$ & M$\downarrow$ & S$\uparrow$ & P$\uparrow$ & M$\downarrow$ & S$\uparrow$ & P$\uparrow$ & M$\downarrow$ & S$\uparrow$ & P$\uparrow$ & M$\downarrow$ & S$\uparrow$ & P$\uparrow$ & M$\downarrow$ \\
   \midrule
  \multirow{2}{*}{P1 (150, 100)} & Lanczos & .646 & 16 & .096 & .776 & 19 & .068 & .789 & 18 & .070 & .870 & 20 & .046 & .776 & 19 & .068 & .768 & 21 & .038 \\
    & Ours & \textbf{.724} & \textbf{19} & \textbf{.071} & \textbf{.912} & \textbf{24} & \textbf{.036} & \textbf{.970} & \textbf{30} & \textbf{.013} & \textbf{.932} & \textbf{25} & \textbf{.025} & \textbf{.912} & \textbf{24} & \textbf{.036} & \textbf{.780} & \textbf{22} & \textbf{.031} \\
  \midrule \midrule
  \multirow{2}{*}{P2 (150, 1500)} & Lanczos & .242 & 10 & .237 & .247 & 10 & .239 & .238 & 10 & .242 & .312 & 12 & .126 & .247 & 10 & .239 & .653 & 17 & .083 \\
    & Ours & \textbf{.398} & \textbf{12} & \textbf{.180} & \textbf{.459} & \textbf{13} & \textbf{.161} & \textbf{.503} & \textbf{12} & \textbf{.177} & \textbf{.523} & \textbf{14} & \textbf{.099} & \textbf{.459} & \textbf{13} & \textbf{.161} & \textbf{.658} & \textbf{19} & \textbf{.063} \\
  \midrule \midrule
  \multirow{2}{*}{P3 (800, 100)} & Lanczos & .303 & 12 & .194 & .292 & 12 & .199 & .422 & 14 & .146 & .211 & 11 & .217 & .292 & 12 & .199 & .367 & 15 & .116 \\
    & Ours & \textbf{.393} & \textbf{15} & \textbf{.131} & \textbf{.376} & \textbf{15} & \textbf{.135} & \textbf{.550} & \textbf{16} & \textbf{.101} & \textbf{.230} & \textbf{13} & \textbf{.156} & \textbf{.376} & \textbf{15} & \textbf{.135} & \textbf{.496} & \textbf{19} & \textbf{.070} \\
  \midrule \midrule
  \multirow{2}{*}{P4 (800, 1500)} & Lanczos & .274 & 12 & .199 & .263 & 12 & .204 & .394 & 14 & .163 & .186 & 11 & .220 & .263 & 12 & .204 & .319 & 15 & .128 \\
    & Ours & \textbf{.329} & \textbf{14} & \textbf{.158} & \textbf{.314} & \textbf{14} & \textbf{.161} & \textbf{.451} & \textbf{14} & \textbf{.141} & \textbf{.189} & \textbf{12} & \textbf{.160} & \textbf{.314} & \textbf{14} & \textbf{.161} & \textbf{.454} & \textbf{18} & \textbf{.078} \\
  \bottomrule
  \end{tabular}
  \caption{SSIM (S),
  PSNR (P),
  MAE (M) between the predicted percept \(\mathbf{y}\) and target \(\mathbf{x}\),
  reported as mean over images. Patients are defined by axon-map parameters \(\rho\) (percept size) and \(\lambda\) (eccentricity).
  The left block varies implant resolution on F-MNIST,
  and the right block varies the dataset for the $28\times28$ implant.
  All results use a $28{\times}28$ simulation grid.
  Best values for each patient and variation are highlighted in bold.}
  \label{tab:combined_comparison}
\end{table*}

\cref{fig:implant_quality_patient_1} illustrates how reconstruction quality varies with implant resolution and encoding method.
The figure shows that the most pronounced visual differences between our method and the downsampling baselines occur at high electrode counts,
where the optimization-based approach can better exploit the increased spatial resolution.
Conversely,
differences between Lanczos and nearest-neighbor downsampling are most relevant at low electrode counts (low resolution),
where the choice of downsampling filter has a stronger impact on the resulting percept.

\section{Conclusion}
\label{sec:conc}
We presented a training-free stimulus encoder that casts prosthetic encoding as a bound-constrained least-squares problem under a patient-specific linearized perceptual model.
We empirically show that the perceptual models are sparse,
and that enforcing bound constraints is essential to obtain feasible,
non-negative stimuli.
By exploiting this sparsity and warm-starting,
projected residual norm steepest descent avoids the memory costs of pseudo-inverses,
and provides an anytime speed--quality trade-off that is compatible with real-time use.
Across four simulated patients,
implant sizes from $15\times15$ to $100\times100$,
and three datasets,
we consistently improve fidelity over classical downsampling,
with the strongest gains on F-MNIST,
and we observe strong patient--dataset interactions.
As our method depends on the assumed perceptual model,
Human-in-the-loop calibration~\cite{Fauvel2022,Granley2023} remains essential and complementary.
Runtime depends on implant configuration and hardware,
and embedded implementations remain to be validated.
Future work will validate subjective preference in simulated patient studies.
We will further extend the framework to time-dependent perceptual models by employing online linearization \cite{Granley2021,Granley2023}.
Beyond epiretinal implants,
we plan to explore application to subretinal perceptual models~\cite{Goldstein2025}.
\label{sec:refs}
\small
\bibliographystyle{IEEEtran}
\bibliography{IEEEabrv,strings,refs}

\newpage
\appendix
\section*{Supplementary Material}
This supplementary material complements the main paper with additional details and visualizations.
We provide a compact overview of the experimental setup,
an ablation study demonstrating the feasibility of the linearization assumption for the investigated perceptual model,
an additional robustness experiment analyzing patient model mismatch following the protocol of~\cite{Granley2022},
and additional results highlighting the sparse structure of the linear perception model.

\section{Experimental details for reproducibility}
This section summarizes the fixed experimental configuration used throughout the paper,
to support reproducibility.
It includes implant parameters (\Cref{tab:implants}),
fixed pulse2percept simulation parameters (\Cref{tab:fixedparams}),
and the simulated patient configurations (\Cref{tab:patients}),
defined by axon-map parameters \(\rho\) and \(\lambda\).

\begin{table}[thb]
    \centering
    \caption{Implant parameters used in all experiments.}
    \label{tab:implants}
    \begin{tabular}{ccc}
    \hline
    Array size & \# electrodes & Spacing (horiz./vert.) (\(\mu\mathrm{m}\)) \\
    \hline
    $15{\times}15$ & 225 & $400/400$ \\
    $28{\times}28$ & 784 & $200/200$ \\
    $100{\times}100$ & 10{,}000 & $100/100$ \\
    \hline
    \end{tabular}
\end{table}

\begin{table}[thb]
    \centering
    \caption{Fixed pulse2percept simulation parameters used in all experiments.}
    \label{tab:fixedparams}
    \begin{tabular}{lc}
    \hline
    Parameter & Value \\
    \hline
    Eye & Right \\
    Implant position (\(\mu\mathrm{m}\)) & $(0,0)$ \\
    Implant rotation & $0^\circ$ \\
    Number of axon segments & 300 \\
    Number of axons & 200 \\
    Stimulus amplitude (model units) & 30.0 \\
    \hline
    \end{tabular}
\end{table}

\begin{table}[htb]
    \centering
    \caption{Simulated patients (axon map parameters) from~\cite{Granley2022}.
    Parameters are given in \si{\micro\metre}.}
    \label{tab:patients}
    \begin{tabular}{lcc}
    \hline
    Patient & \(\rho\) & \(\lambda\) \\
    \hline
    Patient 1 & 150 & 100 \\
    Patient 2 & 150 & 1500 \\
    Patient 3 & 800 & 100 \\
    Patient 4 & 800 & 1500 \\
    \hline
    \end{tabular}
\end{table}

\section{Ablation study: linearization assumption}
\Cref{tab:linear_vs_nonlinear_fmnist} compares method performance on the linearized and nonlinear perceptual models.
For the patients with \(\lambda = 100\),
the linear model assumption is well justified,
as performance is very similar on the linearized and nonlinear evaluation.
For the patients with \(\lambda = 1500\),
the model is less linear,
but the quantitative results remain substantially better than the pseudo-inverse baseline (PINV),
and the Lanczos baseline,
when evaluated on the nonlinear model which does not allow negative amplitudes.
Furthermore,
PINV is not suited for patients with \(\rho = 800\),
even when evaluated on the (unphysical) linearized model which allows negative currents.
Overall,
this supports using the linearized model for optimization,
while reporting performance on the more realistic nonlinear model.

\begin{table}[thb]
    \centering
    \small
    \begin{tabular}{l @{\hspace{5pt}} l @{\hspace{5pt}} c @{\hspace{3pt}} c @{\hspace{3pt}} c @{\hspace{8pt}} c @{\hspace{3pt}} c @{\hspace{3pt}} c @{\hspace{0pt}}}
    \toprule
    \multirow{2}{*}{Patient $(\rho,\lambda)$} & \multirow{2}{*}{Method} & \multicolumn{3}{c}{Linear} & \multicolumn{3}{c}{Nonlinear} \\
    \cmidrule(l{0pt}r{11pt}){3-5}\cmidrule(l{0pt}r{3pt}){6-8}
    & & S$\uparrow$ & P$\uparrow$ & M$\downarrow$ & S$\uparrow$ & P$\uparrow$ & M$\downarrow$ \\
    \midrule
    \multirow{3}{*}{P1 (150, 100)} & Lanczos & $.76$ & $18$ & $.07$ & $.78$ & $19$ & $.07$ \\
      & PINV & $\textbf{1.00}$ & $\textbf{125}$ & $\textbf{.00}$ & $.13$ & $-14$ & $2.76$ \\
      & Ours & $.91$ & $24$ & $.03$ & $\textbf{.91}$ & $\textbf{24}$ & $\textbf{.04}$ \\
    \midrule \midrule
    \multirow{3}{*}{P2 (150, 1500)} & Lanczos & $.39$ & $12$ & $.17$ & $.25$ & $10$ & $.24$ \\
      & PINV & $\textbf{1.00}$ & $\textbf{102}$ & $\textbf{.00}$ & $.11$ & $-19$ & $4.34$ \\
      & Ours & $.71$ & $18$ & $.08$ & $\textbf{.45}$ & $\textbf{13}$ & $\textbf{.16}$ \\
    \midrule \midrule
    \multirow{3}{*}{P3 (800, 100)} & Lanczos & $.29$ & $12$ & $.20$ & $.29$ & $12$ & $.20$ \\
      & PINV & $.00$ & $-25$ & $14.88$ & $.00$ & $-47$ & $194.81$ \\
      & Ours & $\textbf{.36}$ & $\textbf{15}$ & $\textbf{.14}$ & $\textbf{.36}$ & $\textbf{15}$ & $\textbf{.14}$ \\
    \midrule \midrule
    \multirow{3}{*}{P4 (800, 1500)} & Lanczos & $.26$ & $11$ & $.22$ & $.26$ & $12$ & $.20$ \\
      & PINV & $\textbf{.51}$ & $1$ & $.73$ & $.04$ & $-35$ & $32.51$ \\
      & Ours & $.32$ & $\textbf{14}$ & $\textbf{.16}$ & $\textbf{.29}$ & $\textbf{13}$ & $\textbf{.17}$ \\
    \bottomrule
    \end{tabular}
    \caption{SSIM (S),
    PSNR (P),
    and MAE (M) between the predicted percept \(\mathbf{y}\) and target \(\mathbf{x}\),
    reported as mean over images of the F-MNIST dataset.
    Patients are defined by axon-map parameters \(\rho\) (percept size) and \(\lambda\) (eccentricity).
    The left block evaluates methods on the \emph{linearized} model,
    while the right block evaluates methods on the \emph{nonlinear} model (PINV adjusted as in main paper).
    All results use a $28{\times}28$ simulation grid.
    The pseudo-inverse method (PINV) produces significantly worse results on the nonlinear model,
    because the input needs to be made physically feasible before evaluation.
    Best values for each patient and variation are highlighted in bold.}
    \label{tab:linear_vs_nonlinear_fmnist}
  \end{table}

\section{Robustness to patient model mismatch}
This experiment follows the patient model mismatch protocol reported in the supplementary material of~\cite{Granley2022}.
We study the sensitivity of the proposed projected residual norm steepest descent encoder to misspecification of the axon-map parameters,
\(\rho\) and \(\lambda\),
used to construct the linear perception matrix \(\mathbf P_{\phi'}\).
Specifically,
we generate stimuli using misspecified parameter values in \(\mathbf P_{\phi'}\),
while evaluating the resulting percepts under the true patient parameters \(\phi\),
and we report the performance gain relative to Lanczos downsampling.
For a fair comparison,
we apply the same global scaling described in the Experiments section of the main paper to both Lanczos and our method,
and compute this scaling using the true patient parameters.

\begin{figure}[thb]
  \centering
  \includegraphics[width=\columnwidth]{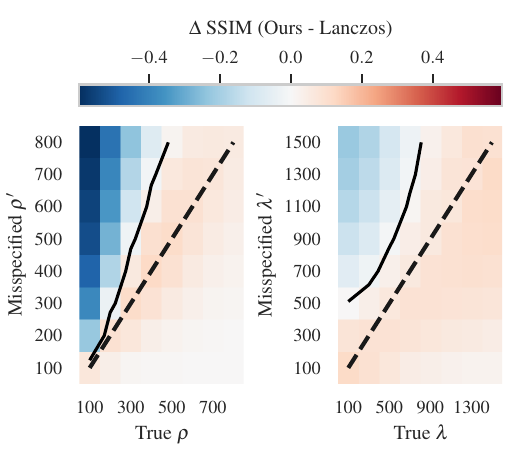}
  \caption{Patient model mismatch experiment,
  inspired by the protocol in the supplementary material of~\cite{Granley2022}.
  The heatmaps show the SSIM difference gain of our method over Lanczos downsampling,
  as a function of misspecified axon-map parameters used during optimization.
  The solid line indicates the zero crossing where our method matches Lanczos.
  The dotted line indicates the gain for \texttt{misspecified=True},
  i.e.,
  when the parameters are estimated correctly.
  Left:
  misspecified \(\rho\) for fixed \(\lambda = 200\).
  Right:
  misspecified \(\lambda\) for fixed \(\rho = 250\).}
  \label{fig:model_mismatch_ssim}
\end{figure}

\Cref{fig:model_mismatch_ssim} shows that we can reproduce the experiment trends reported in~\cite{Granley2022},
indicating a similar performance profile to neural network based approaches under parameter mismatch.
Furthermore,
we observe that setting the misspecified \(\rho\) or \(\lambda\) lower than the true value is less harsh on performance than setting the values too high,
suggesting that conservative parameter choices can be preferable when calibration is uncertain.

\section{Additional experimental results and visualizations showcasing the sparsity of the problem}
\begin{figure}[thb]
  \centering
  \includegraphics[width=\columnwidth]{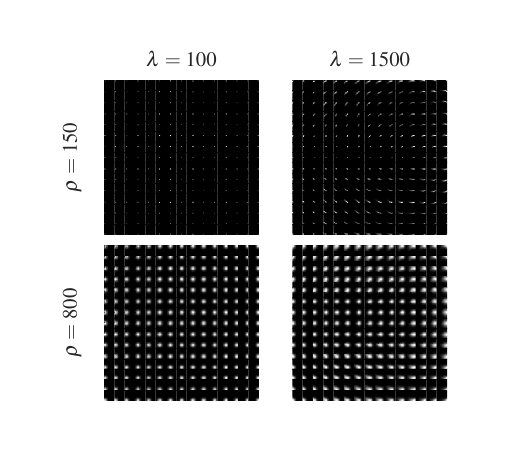}
  \caption{Qualitative visualization of the (sparse) perception matrix $\mathbf P_{\phi} \in \RR^{M \times N}$ for patients 1--4 with a $15{\times}15$ implant.
  Each subfigure shows the receptive-field maps that constitute $\mathbf P_{\phi}$ in the linear model $\mathbf y_{\mathrm{lin}} = \mathbf P_{\phi}\,\mathbf s$,
  where each subimage corresponds to the activation pattern produced by stimulating a single electrode.
  Figure is inspired by~\cite{Fauvel2022}.
  }
  \label{fig:receptive_fields_all}
\end{figure}

\begin{figure}[thb]
  \centering
  \includegraphics[width=\columnwidth]{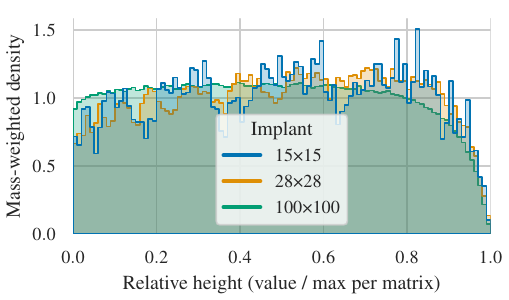}
  \caption{Mass-weighted distribution of entries of the perceptual matrix \(\mathbf P_{\phi}\),
  normalized by \(\max(\mathbf P_{\phi})\),
  for Patient~3 and three different implants.
  Mass-weighted means that each histogram contribution is weighted by the corresponding entry value (relative brightness).
  The plot is representative for the other patients.}
  \label{fig:mass_distribution_patient_3}
\end{figure}

\begin{figure}[thb]
    \centering
    \includegraphics[width=\columnwidth]{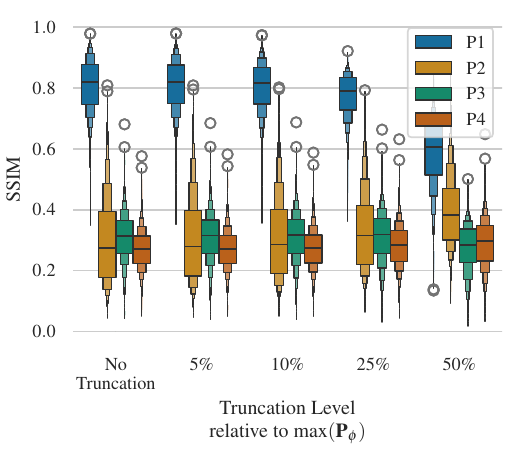}
    \caption{Effect of truncation level when computing the solution on F-MNIST,
    for the $28{\times}28$ implant,
    and a $28{\times}28$ simulated perception grid.
    The horizontal axis shows truncation levels.
    Labels P1--P4 correspond to Patients~1--4 as listed in main paper.
    }
    \label{fig:psnr_truncation_levels}
\end{figure}

The following figures provide additional evaluations and visualizations related to sparsity.
They illustrate how truncating small entries of \(\mathbf P_{\phi}\) affects the computed solution,
and how the electrode-wise activation patterns in \(\mathbf P_{\phi}\) exhibit structured sparsity across patients.
In \Cref{fig:receptive_fields_all},
we provide a qualitative visualization of these electrode-wise activation patterns,
to better illustrate how they look across patients.
In \Cref{fig:mass_distribution_patient_3},
we show a mass-weighted distribution of entries of \(\mathbf P_{\phi}\),
normalized by \(\max(\mathbf P_{\phi})\).
Mass-weighted refers to weighting histogram contributions by the relative brightness (entry value).
The distribution is approximately uniform,
which suggests an underlying density that roughly follows \(1/x\),
and explains why truncating small values can remove many nonzeros,
without removing much total mass.
In \Cref{fig:psnr_truncation_levels},
for truncation levels up to 5\%,
we observe no noticeable change in the computed solution,
consistently across patients.
For Patient~2 and Patient~4,
the improved metrics at higher truncation levels indicate that stronger truncation can in some cases improve quality.

\vfill\pagebreak

\end{document}